%% file: xcache-esnet.tex
\documentclass[sigconf]{acmart}
\AtBeginDocument{%
  \providecommand\BibTeX{{%
    \normalfont B\kern-0.5em{\scshape i\kern-0.25em b}\kern-0.8em\TeX}}}
\usepackage{amsmath,amsfonts}
\usepackage{algorithmic}
\usepackage{graphicx}
\usepackage[caption=false]{subfig}
\usepackage{float}
\usepackage{multirow}
\usepackage{array}
\graphicspath{{figs/}}
\usepackage{url}

\copyrightyear{2021} 
\acmYear{2021} 
\setcopyright{rightsretained} 
\acmConference[SNTA '21]{Proceedings of the 2021 Systems and Network Telemetry and Analytics}{June 21, 2021}{Virtual Event, Sweden}
\acmBooktitle{Proceedings of the 2021 Systems and Network Telemetry and Analytics (SNTA '21), June 21, 2021, Virtual Event, Sweden}\acmDOI{10.1145/3452411.3464441}
\acmISBN{978-1-4503-8386-8/21/06}

\begin{document}
\fancyhead{}
\title{Analyzing scientific data sharing patterns\\for in-network data caching}

\author{Elizabeth Copps}
\affiliation{%
  \institution{Middlebury College}
  \city{Middlebury}
  \state{VT}
  \country{USA}
}
\email{ecopps@middlebury.edu}

\author{Huiyi Zhang}
\affiliation{%
  \institution{University of California, Berkeley}
  \city{Berkeley}
  \state{CA}
  \country{USA}
}
\email{huiyi.zhang@berkeley.edu}

\author{Alex Sim}
\author{Kesheng Wu}
\affiliation{%
  \institution{Lawrence Berkeley Nat'l Laboratory}
  \city{Berkeley}
  \state{CA}
  \country{USA}
}
\email{{asim,kwu}@lbl.gov}

\author{Inder Monga}
\author{Chin Guok}
\affiliation{%
  \institution{Energy Sciences Network}
  \city{Berkeley}
  \state{CA}
  \country{USA}
}
\email{{imonga,chin}@es.net}

\author{Frank W\"{u}rthwein}
\author{Diego Davila}
\author{Edgar Fajardo}
\affiliation{%
  \institution{University of California, San Diego}
  \city{San Diego}
  \state{CA}
  \country{USA}
}
\email{{fkw,didavila,emfajard}@ucsd.edu}

\renewcommand{\shortauthors}{Author, et al.}

\begin{abstract}
The volume of data moving through a network increases with new scientific experiments and simulations. Network bandwidth requirements also increase proportionally to deliver data within a certain time frame.
We observe that a significant portion of the popular dataset is transferred multiple times to different users as well as to the same user for various reasons. 
In-network data caching for the shared data has shown to reduce the redundant data transfers and consequently save network traffic volume.
In addition, overall application performance is expected to improve with in-network caching because access to the locally cached data results in lower latency.
This paper shows how much data was shared over the study period, how much network traffic volume was consequently saved, and how much the temporary in-network caching increased the scientific application performance. It also analyzes data access patterns in applications and the impacts of caching nodes on the regional data repository. 
From the results, we observed that the network bandwidth demand was reduced by nearly a factor of 3 over the study period.

\end{abstract}



\maketitle

\input{intro.tex}
\input{background_hep.tex}
\input{background_esnet.tex}
\input{exp_esnet.tex}
\input{exp_monitoring.tex}

\input{analysis_access.tex}

\input{analysis_traffic.tex}
\input{related.tex}
\input{conc.tex}

\begin{acks}
This work was supported by the Office of Advanced Scientific Computing Research, Office of Science, of the U.S. Department of Energy under Contract No. DE-AC02-05CH11231, and also used resources of the National Energy Research Scientific Computing Center (NERSC). 
This work was also supported by the National Science Foundation through the grants OAC-2030508, OAC-1836650, MPS-1148698 and OAC-1541349.
The authors also gratefully acknowledge Adam Slagell, Anne White, Eli Dart, Eric Pouyoul, George Robb, Goran Pejovic, Kate Robinson, Yatish Kumar, Dima Mishin, Justas Balcas, and Michael Sinatra for their support.
\end{acks}

\bibliographystyle{ACM-Reference-Format}
\bibliography{xcache-esnet}

\end{document}

%% file: intro.tex
\section{Introduction} \label{sec:intro}
The volume of data generated from new scientific experiments and simulations is exponentially increasing, and the data access for such projects drives up the network bandwidth demand and time constraint data delivery requirements. These projects include geographically distributed collaborations, such as the upcoming high-luminosity upgrade of the Large Hadron Collider (LHC) and Large Synoptic Survey Telescope (LSST) experiments. As the research community builds large, one-of-a-kind instruments, the data collected by these instruments are converted into scientifically relevant data sets, which are then used by collaborations between scientists across the world to generate discoveries. As breakthroughs take place in algorithms or hardware and new theories are created, years of older data from scientific instruments might be reprocessed, leading to scientifically significant publications. As this cycle of experiment, discovery, new research, and re-discovery repeats itself, we observed that the popular datasets are delivered multiple times to different users all focused on the same problem. In many cases, the same dataset is delivered multiple times to the same user for various reasons. 

Given the cost of acquiring and maintaining long-term storage, there are typically fewer locations that host data than locations that can compute on the data to produce results. Additionally, the research allocation for large-scale computing can be determined by research allocation or availability of resources in a shared infrastructure. This encourages the same user or multiple users to move data from a few canonical sources to multiple compute locations or the same location, depending on when the compute resource is available to that particular user. Now, if the data source and compute location were to be the same for all users of a particular dataset, sharing data may be possible by deploying large storage on a customer site next to the analysis machines. However, sharing data among geographically distributed users, compute/analysis machines, and data storage centers can only be accommodated with an infrastructure that allows frequent movement of data between these locations. While many solutions have been created to enable efficient movement of data over distances, there is significant time spent by the scientist to ensure that the appropriate data is at the right location before starting the actual analysis. Since this data is typically stored on the disk allocated to the particular scientist, if there are co-located researchers using the same popular dataset for their analysis, multiple copies of that data would be downloaded to each researcher's private storage allocated at the same compute location. However, sharing data among geographically distributed users can be accommodated with some type of content-delivery network.  

One approach to increase data availability is to build an application data cache at the location where the compute resource is available. This local cache is useful in speeding up access to the few popular datasets that might be used by multiple scientists at the same location. However, the benefits may be limited~\cite{stashcache, Fajardo2020} because they depend on the inadvertent intersection of multiple factors, such as scientists with the same interest, computing access, and dataset access at a particular location. 

In-network caching also provides the unique capability for a network provider to design data hotspots into the network topology. The appropriate bandwidth resources and traffic engineering techniques can then be deployed to manage traffic movement and congestion. Even if the shared cache is in the regional resources, accessing remote files may add extra delays to the application performance. These delays would especially be incurred when the files are located deep in the local campus infrastructure. An in-network caching strategy in the middle of the region would reduce the data access latency and increase the overall application performance. For ESnet, which buys transatlantic connectivity to bridge its continental US and European footprints, utilizing in-network caching decreases bandwidth demands on the transatlantic links. In-network caching provides huge cost savings, as subsea bandwidth capacity is significantly more expensive than terrestrial capacity.  This is especially relevant to the High Energy Physics (HEP) community because the LHC instrument is located at CERN in Switzerland, while Tier-1 sites in the US for the ATLAS and CMS experiments are located at Brookhaven National Laboratory in Upton, NY, and Fermi National Accelerator Laboratory in Batavia, IL, respectively.

For this study, we collected data access measurements from the Southern California Petabyte Scale Cache~\cite{socalrepo2018}, where client jobs requested data files for High-Luminosity Large Hadron Collider (HL-LHC) analysis. We studied how much data is shared, how much network traffic volume is consequently saved, and how much the in-network data cache increases application performance. Additionally, we analyzed data access patterns in applications and the impacts of singular data caching nodes to the regional data repository.
From the results, we observed that the network traffic demand for the dataset was reduced by a factor of nearly 3 over the observed period. The data access load is balanced among each node in the regional data cache repository, and the impact of a data cache is evenly distributed to other nodes in the regional federated cache repository. 
Understanding data access patterns and the characteristics of the data access gives us insights into how the data or dataset can be delivered and shared, as well as how the needed resources such as compute, storage and network can be allocated.

%% file: background_hep.tex
\section{Background} \label{sec:background}
\subsection{High Energy Physics (HEP)}
The HEP community has long been one of the largest scientific users of global R\&E networks by volume. Its science depends on globally unique instruments operated by collaborations across hundreds of institutions in dozens of countries. Instruments such as ATLAS and CMS at the LHC in Geneva, Switzerland are comprised of 100 Million electronic channels. They are designed to observe collisions every 25 nanoseconds, i.e. a 40MHz rate of collisions. While complex real-time decision logic, implemented via a mix of custom hardware and software, brings the data rate down substantially, the annual data volumes in 2018 per instrument reached tens of Petabytes. The data volumes are expected to grow by more than an order magnitude by 2028, as a result of detector and collider upgrades for the so-called "High Luminosity LHC" (HL-LHC) science program. This program is expected to last at least ten years, increasing the integrated collision luminosity by at least a factor of 10.  

The social structure of these globally unique science endeavors is such that the required computing and storage infrastructure is funded via global in-kind contributions by all participating countries that can afford to do so. In the past, the share of the single largest national contributor has been roughly 30\%. The global R\&E networks are thus an integral part of the cyberinfrastructure of these science programs.

To prepare for a factor of $\sim$30 increase in annual data volume between 2018 and 2028~\cite{esnetHepReq}, the LHC community is driven towards making any and all data placement much more dynamic. To save costs, data must spend less time on expensive active storage. A conceptual design is being pursued that replicates data between regional "Data Lakes"~\cite{datalakes}, and uses a mixture of remote access and caching within those lakes. 
As Europe dominates the global resource contribution, regional caches that span distances roughly consistent with the geography of major EU countries, i.e. O(1000)km, are attractive.
A regional cache is expected to serve multiple computing centers of varying sizes within that region. This is again driven at least in part by the sociology of in-kind contributions. Leading research universities worldwide host LHC cyberinfrastructure at their institutions, and the regional network POPs thus become a natural location for cache placement. 

\subsection{XCache}
The XRootD software suite~\cite{xrootd2005} provides an ideal architecture for implementing a "Data Lake" as a federated storage infrastructure~\cite{xrootdcms}. It implements tree-architecture, with storage devices as leaves, branch points for data access, and a data discovery protocol that allows for the automatic traversal of the tree, dynamically discovering the physical location of objects or files in the logical namespace. The tree-architecture is ideally suited for the kind of distributed storage infrastructure that the LHC needs for the HL-LHC. XCache~\cite{xcache2014} provides caching functionality as part of the tree. An XCache is its own mini-tree, with a distributed set of cache servers that forms a logical XCache and is connected to a single top level branch. The top level branch is configured to cache a subset of the total federated namespace. Different XCaches can thus serve (partially) (non-)overlapping namespaces. Applications are expected to be directed to a "regional" XCache via the configuration of their runtime environment, e.g. via GeoIP as is done in the OSG Data Federation~\cite{Weitzel2017, stashcache, Fajardo2020}.
Cache misses are handled by XCache as simple XRoot-client calls to the data federation that implements the data lake.
The XRoot data federation is thus the collection of storage systems where data is explicitly placed top-down by "inserting" data into the lake. Then, the role of the XCaches is to provide "low enough" latency data access from the entire compute infrastructure within a region. This allows for compute clusters at smaller institutions that are "stateless" and completely without storage. These clusters are thus less costly to maintain and operate, as long as there is a regional XCache nearby.

In principle, the aforementioned concepts maximize the range of institutions that can provide useful in-kind computing resources to the LHC program, thus maximizing the total size of resources available to the program.

In the present paper we describe the caching behaviour observed in a production pilot XCache system across California, USA. The system has cache servers at the ESnet POP in Sunnyvale, at Caltech, and at UCSD. We call it a "production pilot" because the system was used by the CMS collaboration as part of the Caltech and UCSD Tier-2 center production infrastructure, but it has failovers in place such that we could take down the pilot at anytime. The cache utilization and data access patterns measured are thus those of real CMS data analysis for actual science publications, i.e. work done by physicists who are blissfully unaware that they were part of a cyberinfrastructure experiment.   

%% file: background_esnet.tex
\subsection{Energy Sciences Network (ESnet)} \label{sec:esnet}
The Energy Sciences Network (ESnet) is the US Dept of Energy (DOE) Office of Science’s high-performance network user facility, delivering highly-reliable data transport capabilities optimized for the requirements of large-scale science.  ESnet is stewarded by the Advanced Scientific Computing Research Program (ASCR) and managed and operated by the Scientific Networking Division at Lawrence Berkeley National Laboratory (LBNL). ESnet acts as the primary data circulatory system for science by interconnecting the DOE’s national laboratory system, dozens of other DOE sites, and ~150 research and commercial networks around the world. It allows tens of thousands of scientists at DOE laboratories and academic institutions across the country to transfer vast data streams and access remote research resources in real-time.  

ESnet exists to provide the specialized networking infrastructure and services required by the national laboratories, large science collaborations, DOE user facilities, and the DOE research community.  All together, ESnet provides a foundation for the  nation’s scientists to collaborate on some of the world's most important scientific challenges, including energy, biosciences, materials, and the origins of the universe. Science data traffic across its network has grown at around  60\%\ each year, and traffic has exceeded an exabyte per year since 2019.

%% file: exp_esnet.tex
\section{Experimental Setup} \label{sec:exp}
\subsection{Dataset}
Between May 2020 and Oct. 2020, measurements were collected from the Southern California Petabyte Scale Cache repository. The repository consists of 11 nodes at UCSD with 24TB of storage and 10Gbps network connectivity each, 2 nodes at Caltech with 180TB of storage and 40Gbps network connectivity each, and 1 node at ESnet with 44TB of storage and 40Gbps network connectivity.
The ESnet node is connected at Sunnyvale, CA and has about 10ms of round trip time (RTT) from the rest of the regional cache.
A single node study with the ESnet node is done for the full data collection period (May - Oct. 2020), while the cache utilization and data access pattern study with all 14 nodes in the regional cache is done for the months of June, July and August 2020. Due to monitoring issues from late Aug. 2020 to Oct. 2020, these are the only three months of data available for the study.

%% file: exp_monitoring.tex
\subsection{Monitoring data path}
Every node in the cache is configured (configuration details in~\cite{xrootd-monitoring-manual-50,xrootd-monitoring-manual-51}) to periodically send information about its data accesses via UDP packets. An external collector aggregates these data and eventually sends it to a database at CERN. A single collector at UCSD was used for the entire XCache infrastructure across ESnet, Caltech, and UCSD, and we performed an independent check to ensure that record loss was negligible.
When a user reads a given file, 3 types of events are sent to the collector: 1. a \textit{file open}, 2. one or more \textit{read bytes} requests, and 3. a \textit{file close}. Each one of these events has a unique identifier so the collector can relate them all later on. When a \textit{file close} event is received, the collector builds up a single monitoring record that includes the name of file that was read, information about the client (host and domain name), and the total number of bytes read, among other details.
Using the data analytics infrastructure at CERN, we can then extract the monitoring records described above and obtain historical data of the data accesses on the cache.

%% file: analysis_access.tex
\section{Analysis} \label{sec:analysis}
\subsection{Cache utilization and access patterns}
The cache utilization study may show a few characteristics on data management, data sharing, and data access patterns.

\begin{table}[h!]
\scriptsize
\centering
\caption{
Summary statistics for data accesses at the regional caches from June to Aug. 2020
}
\begin{tabular}{|>{\centering}p{1.2cm}||>{\centering}p{1.3cm}|>{\centering}p{1.3cm}|>{\centering}p{1.3cm}|>{\centering}p{1.3cm}|} \hline
  & \# of accesses & data transfer size (GB) & shared data size (GB) & Percentage of shared data size \tabularnewline \hline \hline
June 2020 & 1,804,697 & 532,037.7 & 818,956.9 & 60.62\% \tabularnewline \hline
July 2020 & 1,426,585 & 354,452.8 & 764,351.3 & 68.32\% \tabularnewline \hline
Aug 2020 & 995,324 & 249,583.5 & 586,188.8 & 70.14\% \tabularnewline \hline
Total &  4,226,606 & 1,136,074.0 & 2,169,497.0 & 65.63\% \tabularnewline \hline
Daily average & 48,029.61 & 12,909.93 & 24,653.37 &  \tabularnewline \hline
\end{tabular}
\label{tab:summary_data_all}
\end{table}

Table \ref{tab:summary_data_all} shows the basic statistics on the data access activities for all caching nodes during the study period (from June to Aug. 2020). 
The "data transfer size" column in Table \ref{tab:summary_data_all} indicates the data volume resulting from the "cache misses", when a data file was accessed for the first time. For cache misses, the caching nodes in the region did not have the data, resulting in a data transfer from the remote site to one of the local caching nodes. 
The "shared data access size" column refers to the data volume from the "cache hits", when the data file was already in the cache and readily available for the application to access. 
The shared data accesses count is for repeated accesses only and correspond to the network traffic savings.
The "Percentage of shared data size" column indicates the percentage of the total data access size that was shared rather than transferred. This percentage is closely related to the network traffic demand reduction. As the table shows, it is expected that the percentage of shared data size increases with time. The percentage would stop increasing at a certain level because the cache can only hold a set amount of the popular dataset. 

\begin{figure}[htb!]
\centering
\includegraphics[width=\columnwidth]{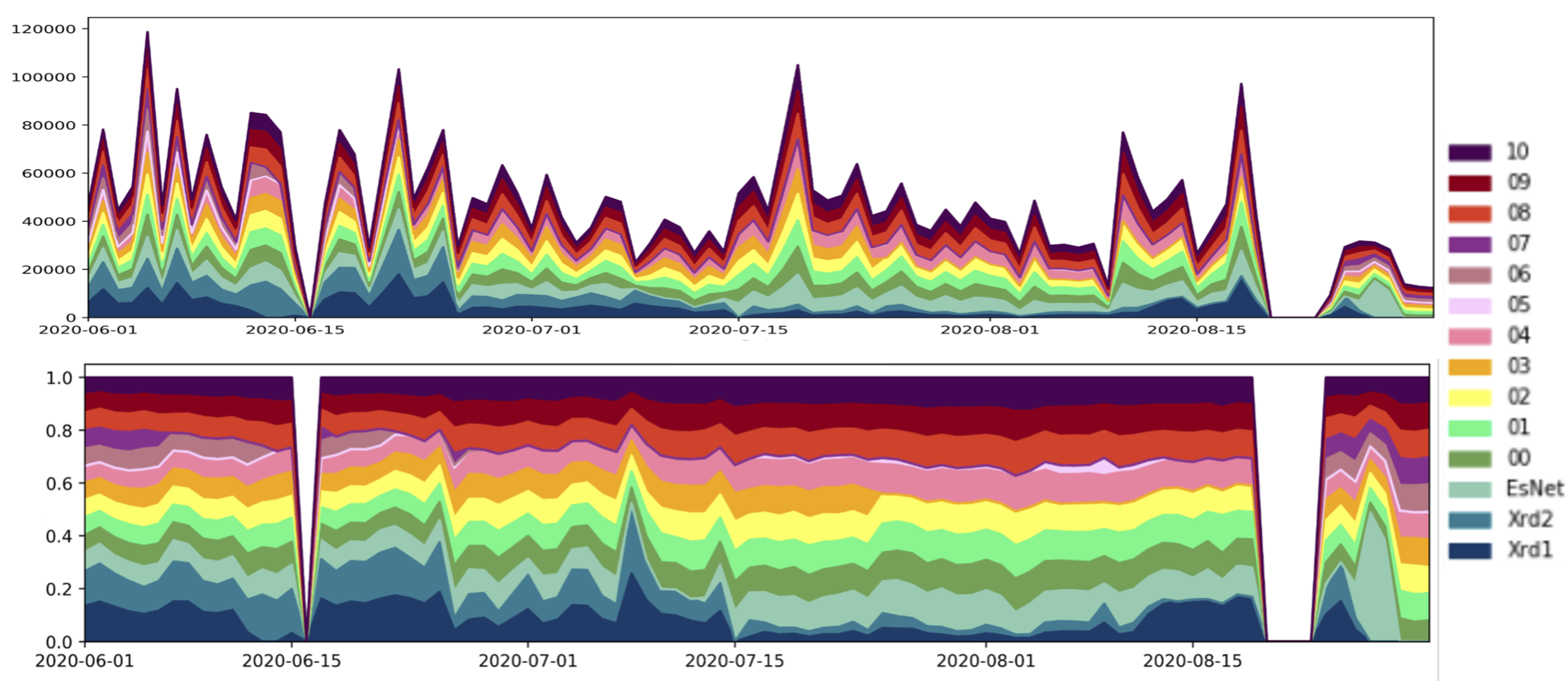}
\caption{
    Daily total data access counts and proportion of total accesses on each node in the regional cache
}
\label{fig_access_all}
\end{figure}

Figure \ref{fig_access_all} shows the daily total data access counts, combining cache miss counts (i.e. data transfers) and cache hit counts (i.e. shared data accesses). Two larger cache nodes have the cache sizes of 7.5 times of the most smaller cache nodes, but the accesses do not cover 7.5 times higher than the smaller caches, nor the shared data ratio. It indicates that there is a limited relationship between the cache size and the data access activities, and the access policy at the regional gateway may also have an influence on the data access activities on each node in the region. 
The figure also shows a few down-times of the regional cache and individual cache nodes. In the rest of the regional cache, the down-time of an individual cache node causes an 54 fewer shared data accesses per hour on average and 105 more data transfers per hour on average. Further study could analyze how to minimize the impact of the node down-time on the regional cache.

\begin{figure}[htb!]
\centering
\includegraphics[width=\columnwidth]{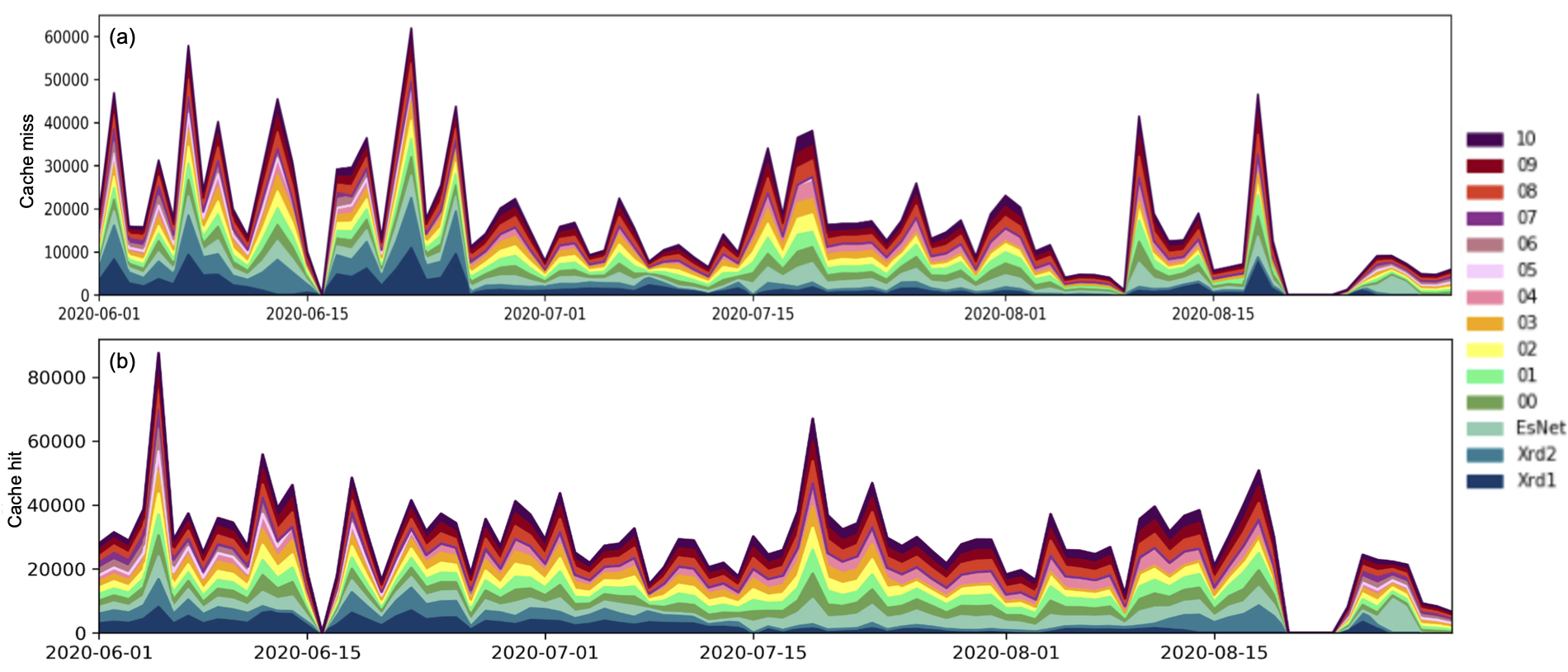}
\caption{
    (a) daily cache miss counts and (b) cache hit counts on each node in the regional cache
}
\label{fig_cachemiss_all}
\end{figure}

\begin{figure}[htb!]
\centering
\includegraphics[width=\columnwidth]{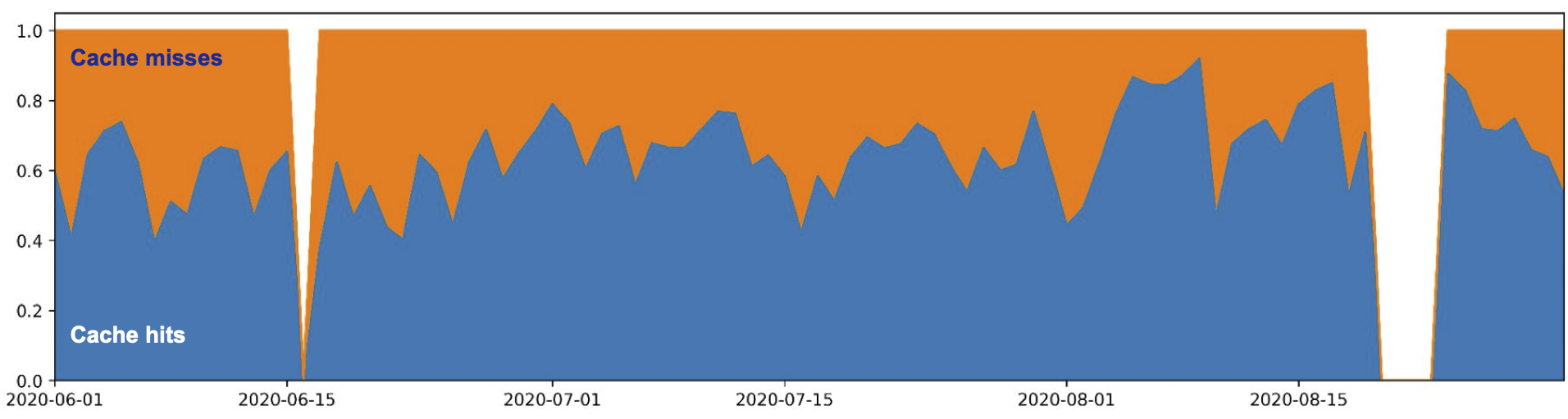}
\caption{
    Daily proportion of cache miss counts (orange area) and cache hit counts (blue area) in the regional cache
}
\label{fig_cache_miss_hit_prop_all}
\end{figure}

Figure \ref{fig_cachemiss_all} shows the daily cache miss counts and cache hit counts on each node in the regional cache that were separated from the total data access counts in the Figure \ref{fig_access_all}. There is a similar proportional distribution of the counts on each node for both cache misses and cache hits. The ratio of the cache hit counts proportional to the cache misses gets higher as times goes on, shown in Figure \ref{fig_cache_miss_hit_prop_all}. 

\begin{figure}[htb!]
\centering 
\subfloat[Daily]{%
  \includegraphics[width=0.5\columnwidth]{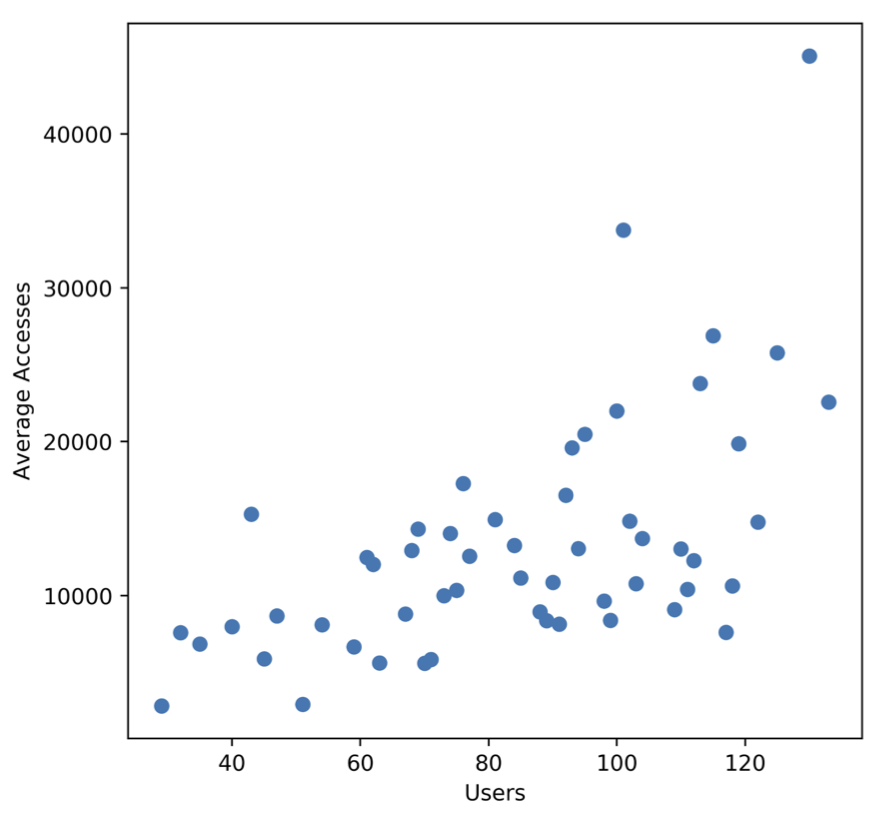}
  \label{fig_daily_users_ave_access}
}
\subfloat[\textit{Hourly}]{%
  \includegraphics[width=0.5\columnwidth]{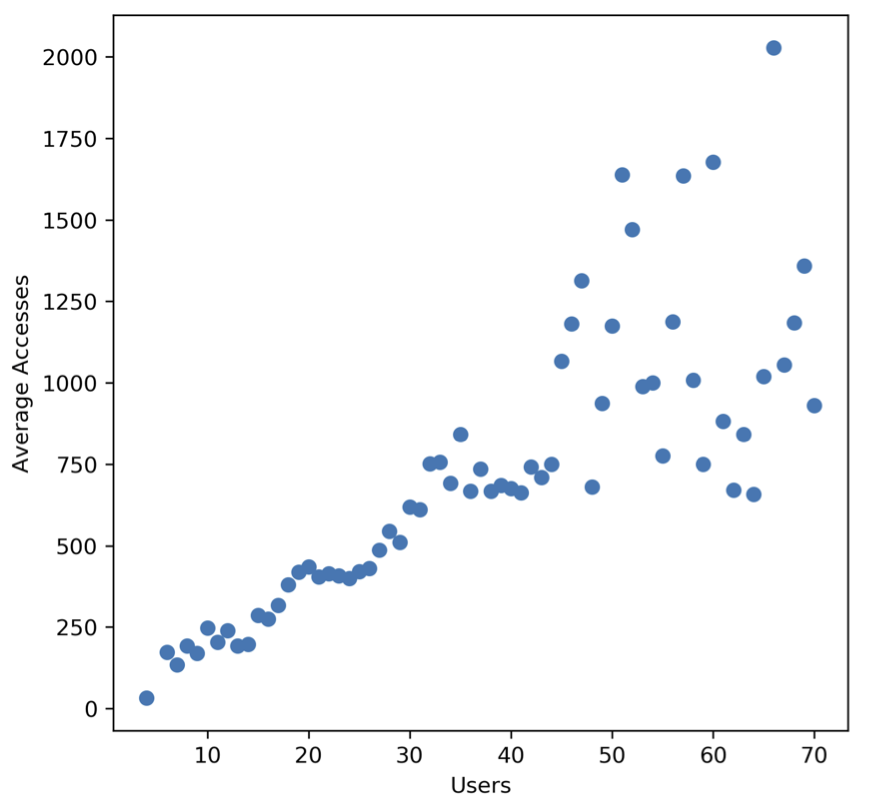}
  \label{fig_hourly_users_ave_access}
}\newline
\caption{Distribution of the number of users and the average data access counts}
\label{fig_user_access_all}
\end{figure}

Figure \ref{fig_user_access_all} shows the distribution of the number of users and the average total data access counts. The hourly distribution, Figure \ref{fig_hourly_users_ave_access}, shows a clear linear relationship between the number of users and the average total data access counts ($\text{average total data access} = 20.33 * \text{number of users} - 24.09$), and it indicates that the number of total data accesses would be roughly doubled when the number of users is doubled in the regional cache.

\begin{figure}[htb!]
\centering 
\subfloat[\# of users vs. \# of data transfers]{%
  \includegraphics[width=0.5\columnwidth]{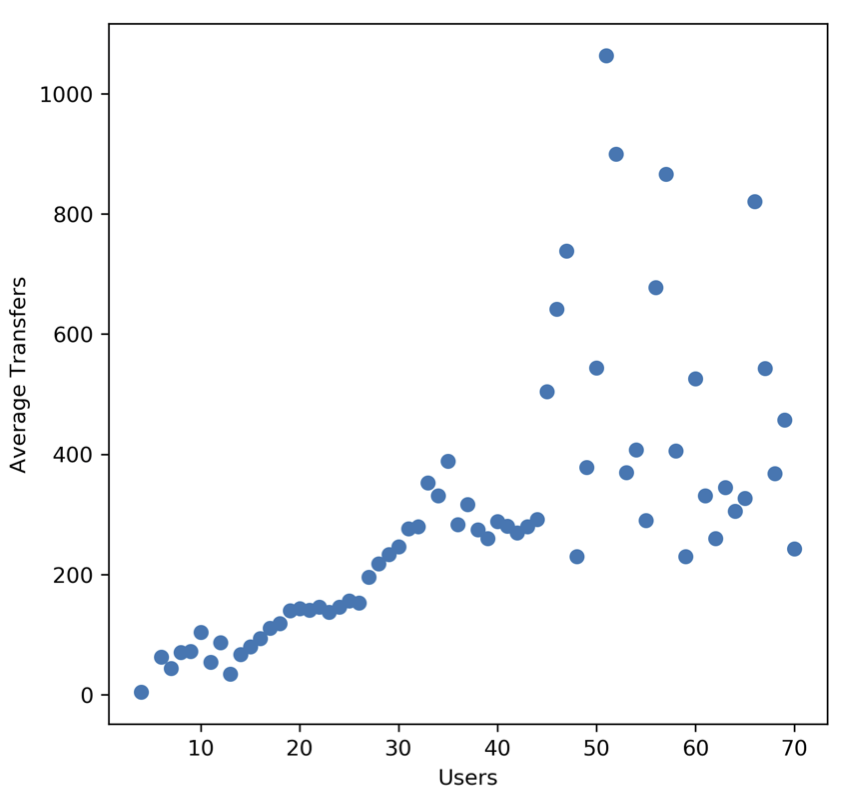}
  \label{fig_hourly_users_ave_transfers}
}
\subfloat[\textit{\# of users vs. \# of shared accesses}]{%
  \includegraphics[width=0.5\columnwidth]{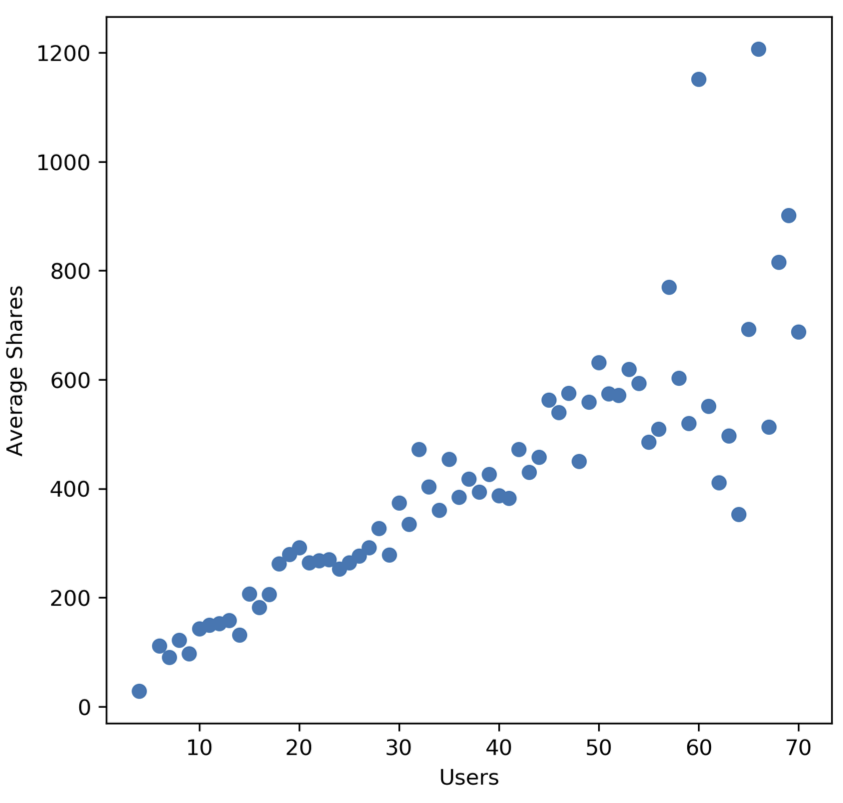}
  \label{fig_hourly_users_ave_shares}
}\newline
\subfloat[\# of accesses vs. data transfer size]{%
  \includegraphics[width=0.5\columnwidth]{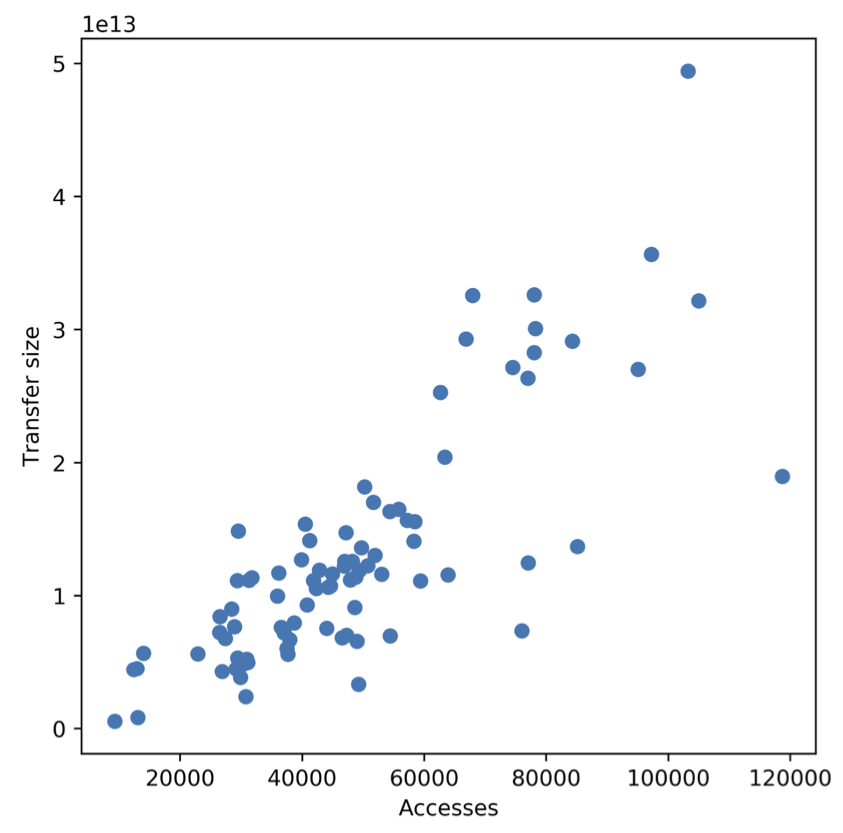}
  \label{fig_hourly_access_transfer_size}
}
\subfloat[\textit{\# of accesses vs. shared data size}]{%
  \includegraphics[width=0.5\columnwidth]{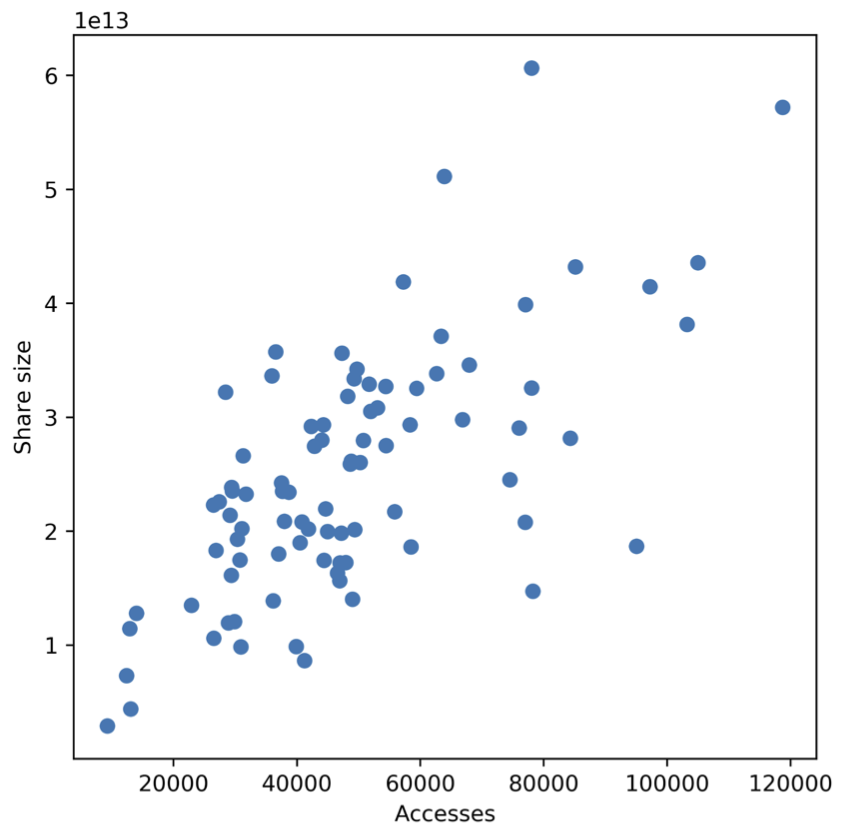}
  \label{fig_hourly_access_shared_size}
}\newline
\caption{Distribution of the number of users, number of data transfers, number of shared data accesses, hourly data transfer volume, and hourly shared data volume}
\label{fig_hourly_size_all}
\end{figure}

Figures \ref{fig_hourly_users_ave_transfers} and \ref{fig_hourly_users_ave_shares} show the distribution of the hourly average data transfer counts and hourly average shared data access counts in terms of the number of users. 
A linear relationship can be found: 
\begin{multline}
\text{average hourly data transfer counts} = \\
10.39 * \text{number of users - 68.4}
\end{multline}
\begin{multline}
\text{average hourly shared data access counts} = \\
10.6 * \text{number of users +21.48}
\end{multline}
The hourly number of the data transfers would be slightly more than double when the number of users is doubled, and the hourly number of the shared data accesses would be estimated slightly less than double when the number of users is doubled.
Figures \ref{fig_hourly_access_transfer_size} and \ref{fig_hourly_access_shared_size} show the distribution of the hourly average data transfer volume and the hourly average shared data volume in terms of the number of data accesses. The figures show that the hourly shared data volume increases more than the hourly data transfer volume when the number of data accesses increases. The increase rate would be expected to be maintained up to certain level where the proportion of the shared data volume is stabilized in the cache utilization.  

\begin{table}[htb!]
\scriptsize
\centering
\caption{
Summary statistics for data accesses at the ESnet cache node from May to Oct. 2020
}
\begin{tabular}{|>{\centering}p{1.2cm}||>{\centering}p{1.3cm}|>{\centering}p{1.3cm}|>{\centering}p{1.3cm}|>{\centering}p{1.3cm}|} \hline
  & \# of accesses & data transfer size (GB) & shared data size (GB) & Percentage of shared data size \tabularnewline \hline \hline
May 2020 & 189,984 & 30,150.50 & 47,986.56 & 61.41\% \tabularnewline \hline
June 2020 & 215,452 & 40,835.23 & 55,929.47 & 57.80\% \tabularnewline \hline
July 2020 & 205,478 & 33,399.81 & 66,457.3 & 66.55\% \tabularnewline \hline
Aug 2020 & 203,806 & 30,819.80 & 68,723.19 & 69.04\% \tabularnewline \hline
Sep 2020 & 165,910 & 10,153.97 & 38,036.19 & 78.93\% \tabularnewline \hline
Oct 2020 & 306,118 & 22,723.93 & 45,614.91 & 66.75\% \tabularnewline \hline  \hline
Total & 1,286,748 & 168,083.27 & 322,747.67 & 65.76\% \tabularnewline \hline
Daily average & 9,674.79 & 1,263.8 & 2,426.67 & \tabularnewline \hline
\end{tabular}
\label{tab:esnet_summary_data}
\end{table}

The cache utilization and data access patterns are further studied for an individual cache node in the regional cache. 
Table \ref{tab:esnet_summary_data} shows the data access activity counts on the ESnet node during the study period (from May to Oct. 2020). It shows that the percentage of the data volume that is shared on the ESnet cache node roughly corresponds to the percentage in the regional cache.

\begin{figure}[htb!]
\centering
\includegraphics[width=\columnwidth]{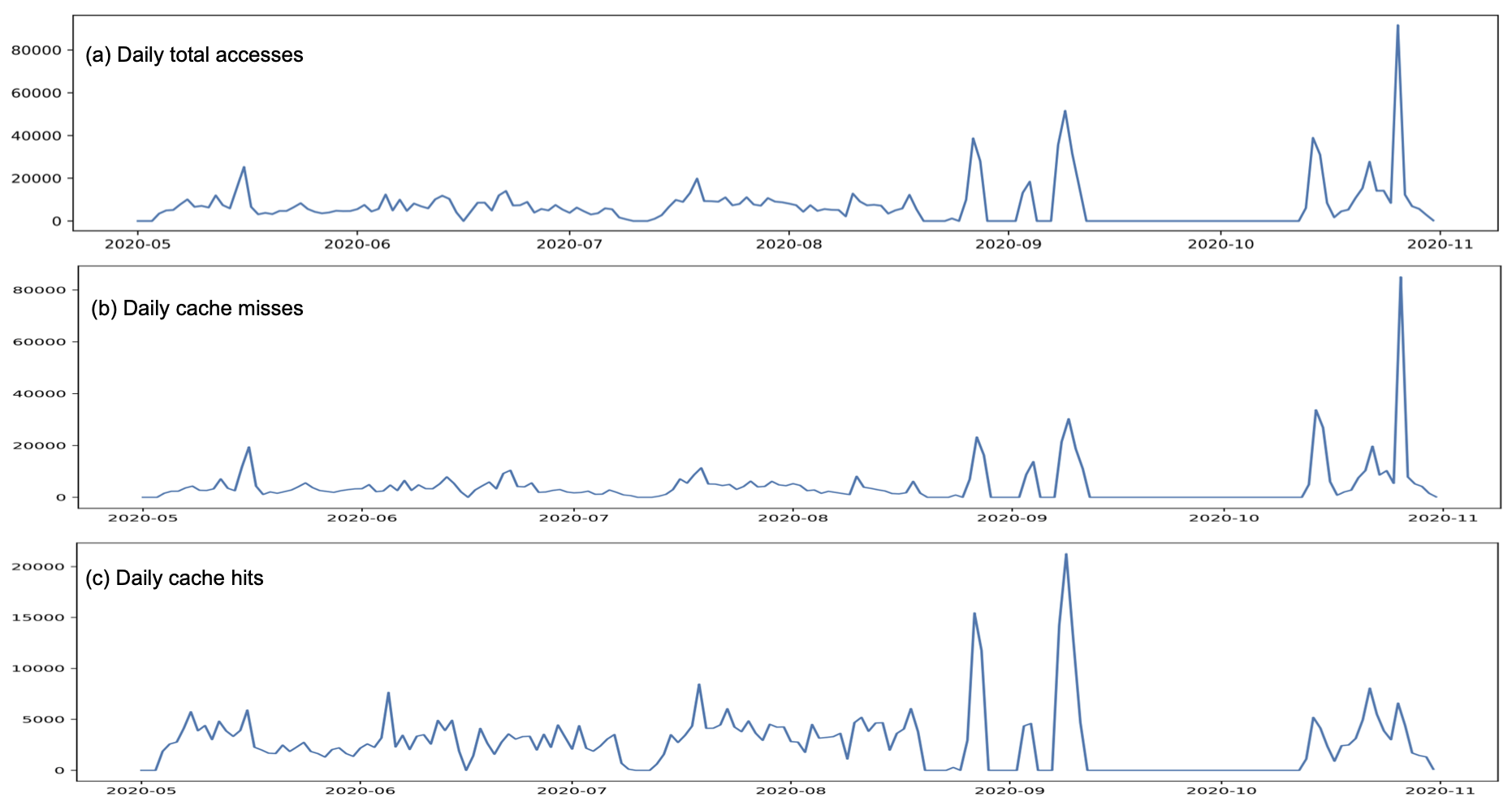}
\caption{
    Number of the daily data access counts on ESnet cache node
}
\label{fig_esnet_access_count}
\end{figure}

Figure \ref{fig_esnet_access_count}a shows the number of daily total data accesses on the ESnet cache node. 
Figure \ref{fig_esnet_access_count}b shows the number of daily data transfers on the ESnet node which corresponds to the cache misses and the first time accesses. 
Figure \ref{fig_esnet_access_count}c shows the number of daily shared data accesses on the ESnet node which corresponds to the cache hits and the repeated accesses.
They all show similar data access patterns to the regional cache, indicating that the regional cache gateway runs the same policy for all cache node in the region, unless an anomalous behaviour occurs. 
As described earlier, a few days in Aug. Sep. and Oct. show down-times, but they are due to the monitoring issues. The analysis does not include those days without collected measurements.
Figure \ref{fig_esnet_access_count}b shows a spike in the number of data transfers towards the end of Oct. which is reflected in the total number of data accesses.

\begin{figure}[htb!]
\centering
\includegraphics[width=\columnwidth]{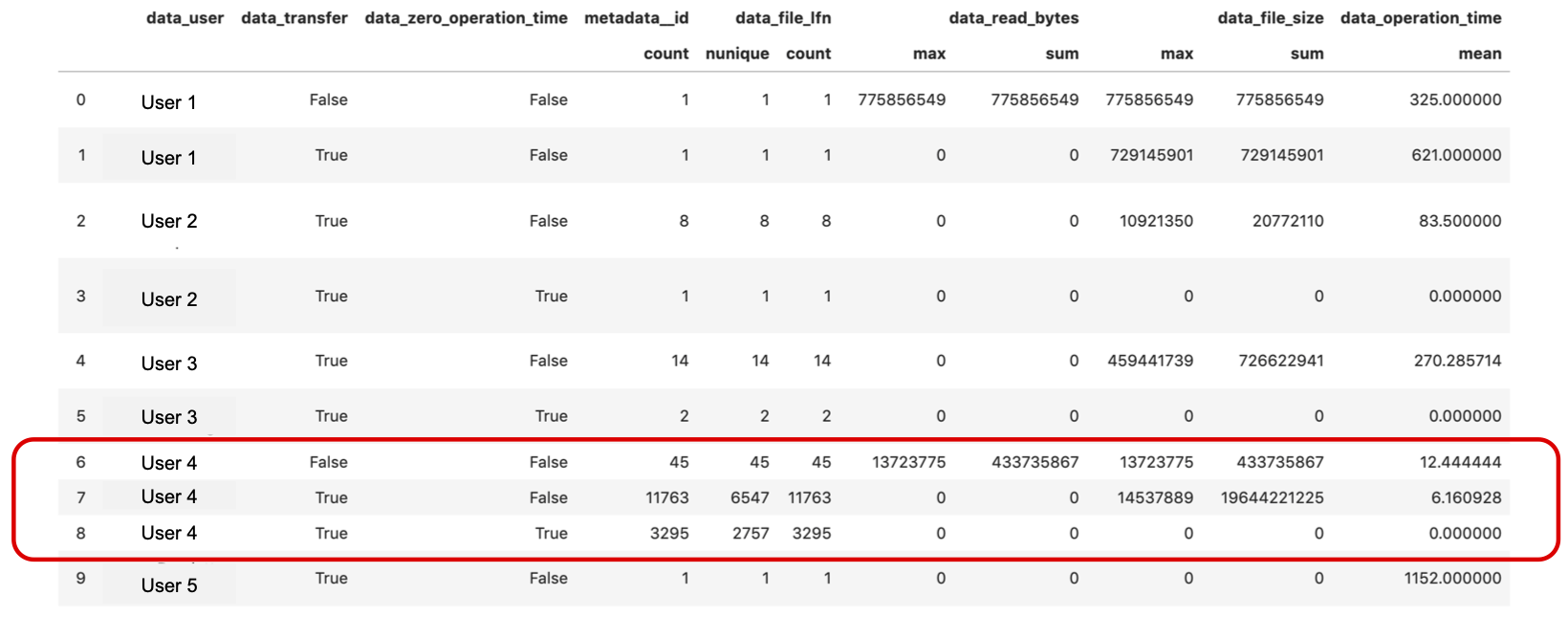}
\caption{
    Data transfer spike in Oct. 26, 2020 on ESnet cache node
}
\label{fig_transfer_spike_esnet}
\end{figure}

\begin{figure}[htb!]
\centering
\includegraphics[width=\columnwidth]{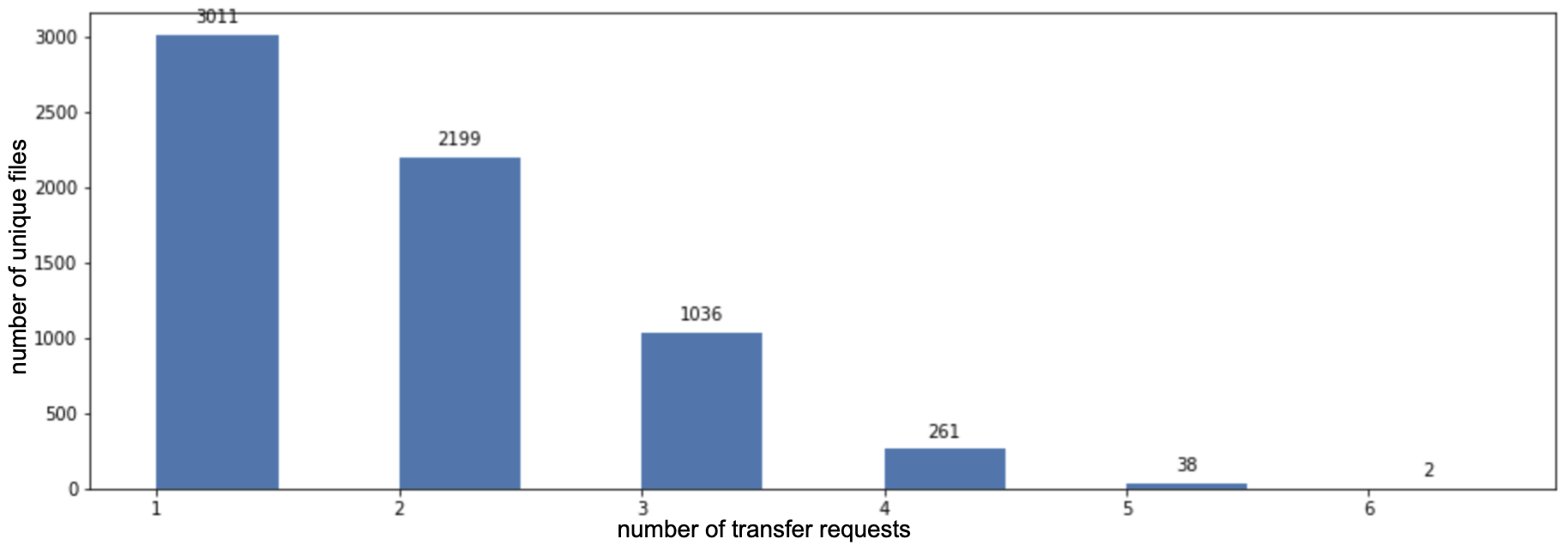}
\caption{
    Number of repeated data transfers for ESnet cache node in Oct. 26, 2020
}
\label{fig_transfer_spike_hist_esnet}
\end{figure}

As shown in Figure \ref{fig_transfer_spike_esnet}, the data transfer records on the ESnet node between 12pm and 1pm in Oct. 26, 2020 show 15,058 transfer operations for one user. 
While 3,295 transfers have 0 file transfer size in 0 second, the other 11,763 transfers moved 6,547 unique files in total 19.64GB. 
Figure \ref{fig_transfer_spike_hist_esnet} shows how many files of the 6,547 unique files have been transferred multiple times.
5,216 transfers were repeated for reasons beyond the scope of this study. This kind of anomalous behavior can be prevented by future policies in the caching service.

\begin{figure}[htb!]
\centering
\includegraphics[width=\columnwidth]{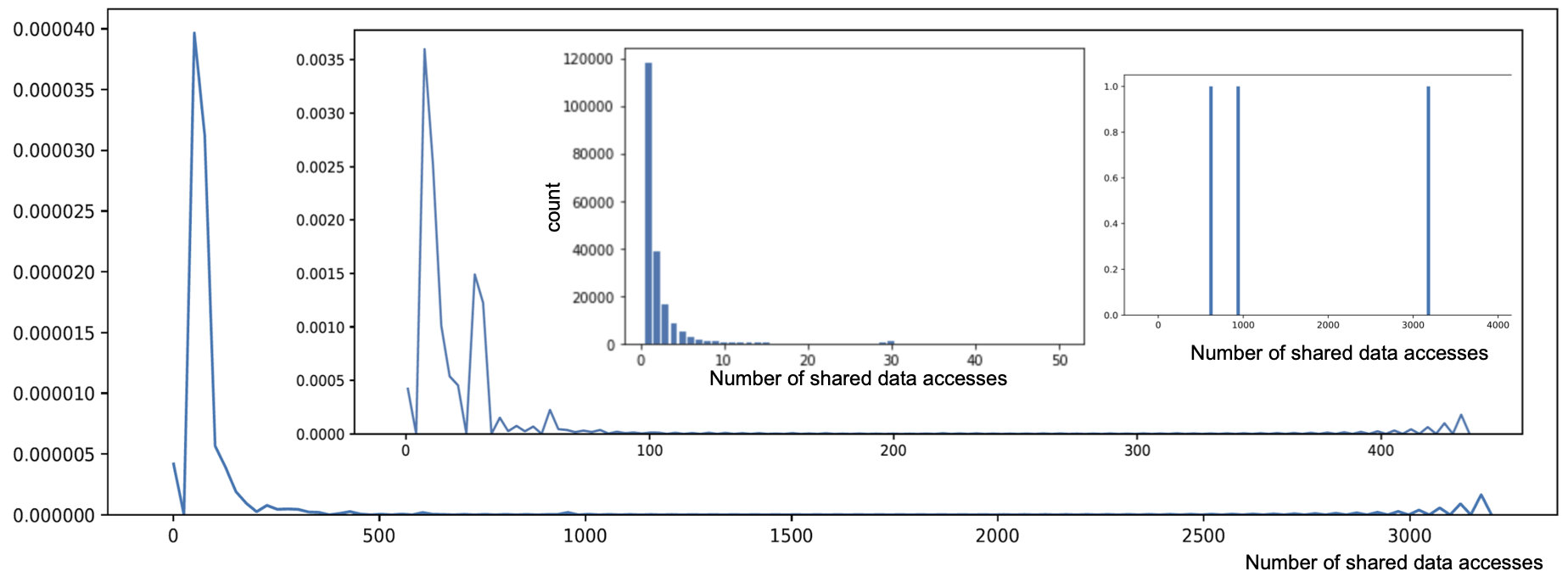}
\caption{
    Distribution of the number of shared data accesses on ESnet cache node
}
\label{fig_dist_shared_esnet}
\end{figure}

Figure \ref{fig_dist_shared_esnet} shows the distribution of the number of shared data accesses (repeated data accesses) for the ESnet node during the study period. 
The total number of data accesses, including the first access and repeated accesses, was 1,286,748 during the measured period.
There were a total of 490,944 shared data accesses for 198,940 unique files. 
While the majority of the data files was accessed only once, 3 files were accessed 4,762 times. 

\begin{figure}[htb!]
\centering
\includegraphics[width=\columnwidth]{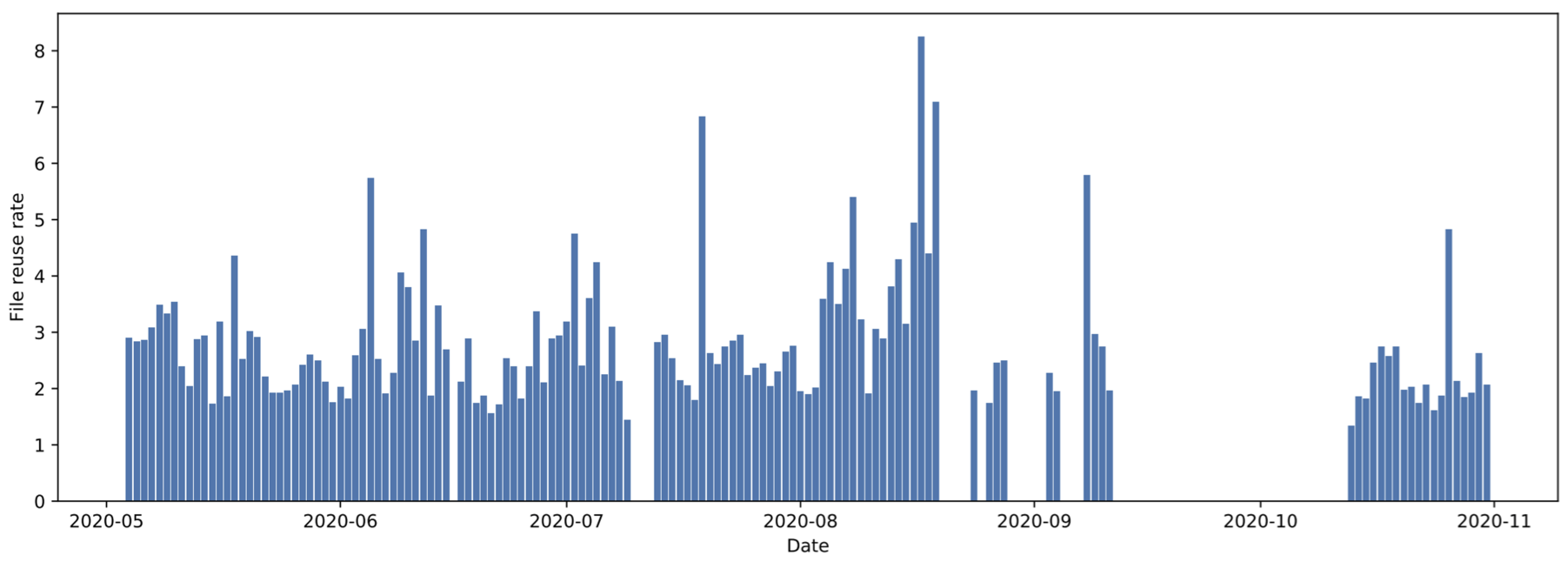}
\caption{
    Daily data file reuse rates on ESnet cache node
}
\label{fig_reuse_rates_esnet}
\end{figure}

Figure \ref{fig_reuse_rates_esnet} shows the daily file reuse rates on the ESnet node, which is calculated by eqn. (~\ref{eqn_file_reuse_rate}).
\begin{multline}
\text{file reuse rate} = \\
\frac{\text{shared data access counts}}{\text{number of unique files}} =  \frac{\text{cache hits}}{\text{number of unique files}}
\label{eqn_file_reuse_rate}
\end{multline}
Note that the daily file reuse rates do not depend on the cache misses and indicate how many unique files are included in the cache hits.
The file reuse rate is closely related to the network traffic savings.

%% file: analysis_traffic.tex
\subsection{Network utilization}
The cache utilization study may also show a few characteristics of the network usage, where shared data directly contributes to the network traffic savings. 

\begin{figure}[htb!]
\centering
\includegraphics[width=\columnwidth]{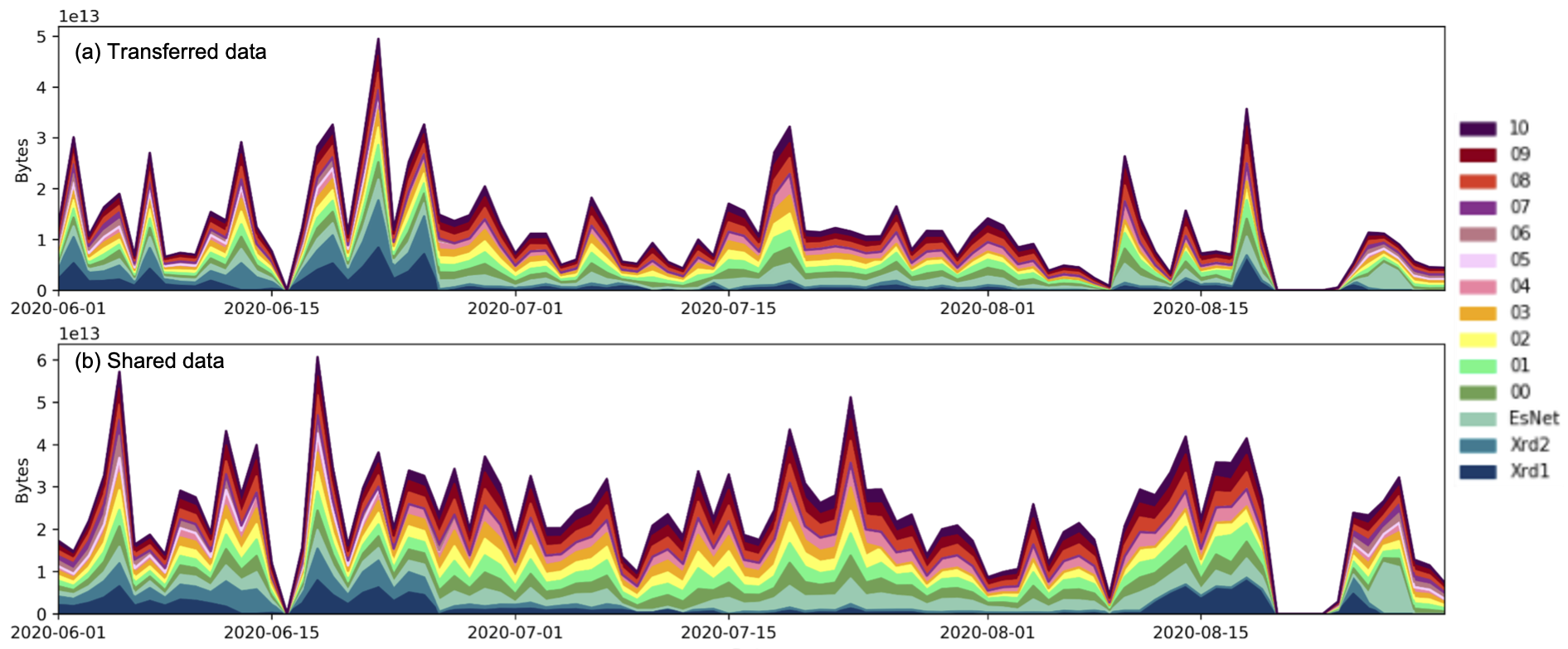}
\caption{
    Daily data transfer size and shared data size on each node in the regional cache
}
\label{fig_transfer_shared_size_all}
\end{figure}

Figure \ref{fig_transfer_shared_size_all}a shows the volume of the daily data transfers, and Figure \ref{fig_transfer_shared_size_all}b shows the daily shared data size on each node in the regional cache. They follow roughly similar patterns to the daily number of data transfer counts and shared data access counts in the regional cache.
Larger caches again do not take a proportionally larger load in the shared data volume, the same as with the data access counts. It would be a further study point to optimize the cache size for the data sharing and network traffic savings.

\begin{figure}[htb!]
\centering
\includegraphics[width=\columnwidth]{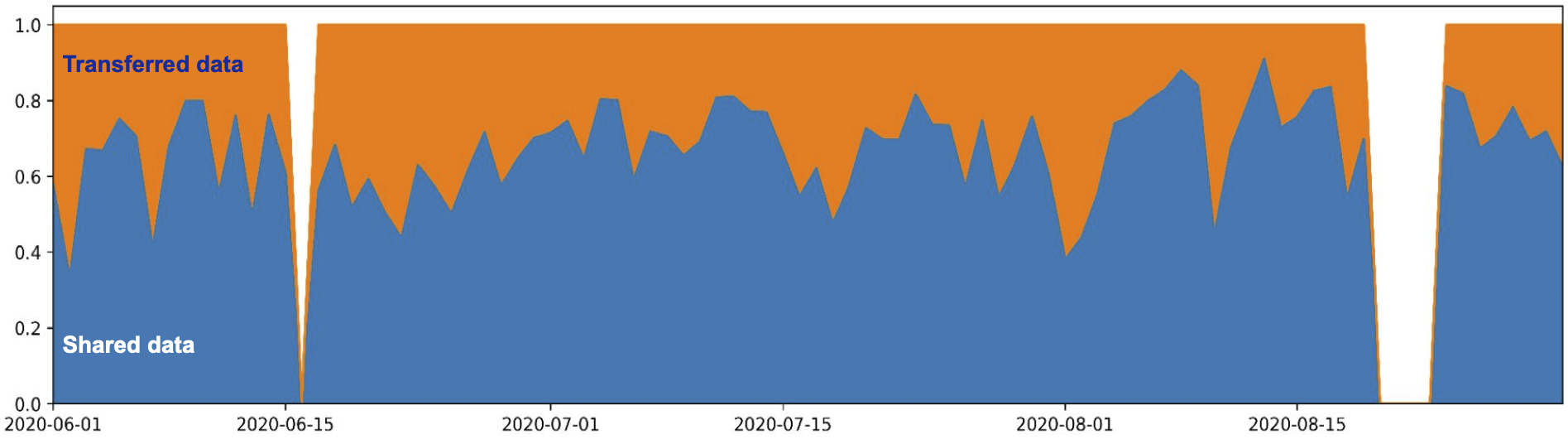}
\caption{
    Daily proportion of data transfer size (orange area) and shared data size (blue area) in the regional cache
}
\label{fig_transfer_shared_prop_all}
\end{figure}

As shown in Figure \ref{fig_cache_miss_hit_prop_all}, the proportion of the daily number of shared data accesses increases as time goes on, and Figure \ref{fig_transfer_shared_prop_all} shows a similar pattern for the volume of the daily shared data in the regional cache. The ratio of the volume of the data transfers to the volume of the shared data also increases as time goes on.

\begin{figure}[htb!]
\centering
\includegraphics[width=\columnwidth]{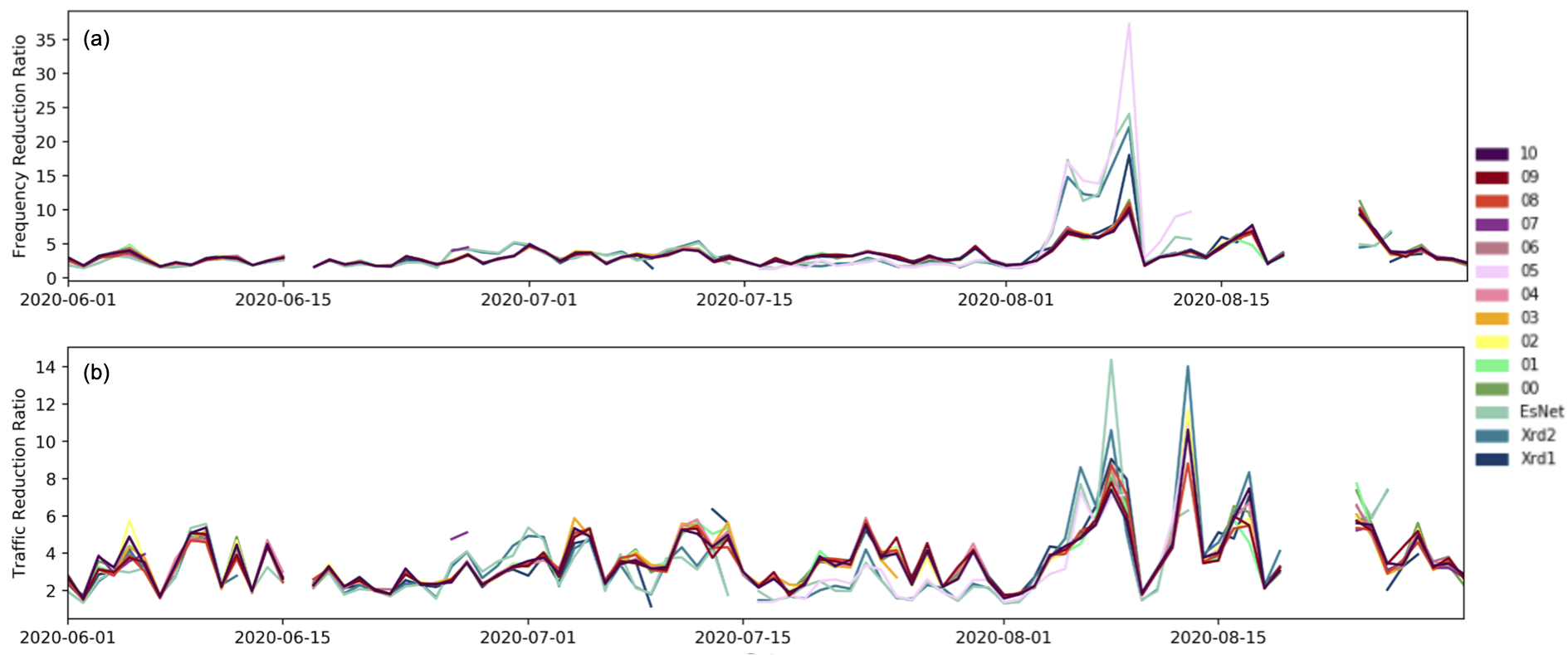}
\caption{
    Network demand reduction rate for each node in the regional cache; (a) data transfer frequency demand, (b) traffic demand
}
\label{fig_net_reduction_ratio_each}
\end{figure}

\begin{figure}[htb!]
\centering
\includegraphics[width=\columnwidth]{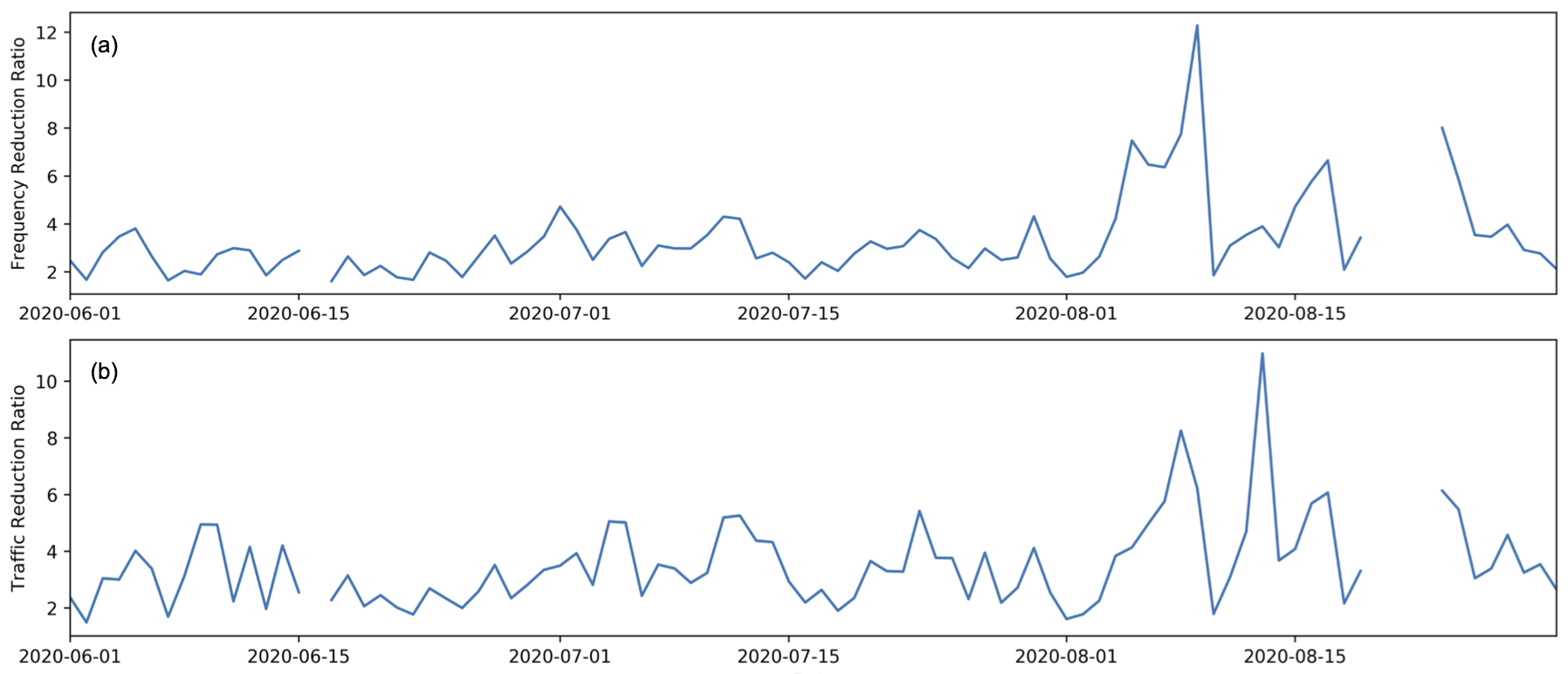}
\caption{
    Network demand reduction rate in the regional cache; (a) data transfer frequency demand, (b) traffic demand
}
\label{fig_net_reduction_ratio}
\end{figure}

We have studied network utilization in two aspects: frequency, indicating how many times data transfers occur regardless of the data transfer size, and traffic volume, indicating how much data is moved from remote sites to the local cache.
Figure \ref{fig_net_reduction_ratio_each} shows the daily network demand reduction rates for each node in the regional cache. 
The network transfer frequency demand reduction rates, calculated by the eqn. (\ref{eqn_freq_reduction}), are shown in Figure \ref{fig_net_reduction_ratio_each}a. Each node shows similar patterns to each other except for a few days in Aug. where a few nodes in the regional cache show significantly higher transfer frequency demand reduction rates. 
Those few days correspond to the higher proportion of the shared data access counts in early Aug. in Figure \ref{fig_cache_miss_hit_prop_all}a. 
The network traffic demand reduction rates, calculated by the eqn. (\ref{eqn_traffic_reduction}) and shown in Figure \ref{fig_net_reduction_ratio_each}b also follow a similar pattern for those few days in early August where Figure \ref{fig_transfer_shared_size_all}a shows significantly higher traffic demand reduction rates.

\vspace{-0.4cm}
\begin{multline}
\text{network transfer frequency demand reduction rate} = \\
\frac{\text{(total shared access counts + total transfer counts)}}{\text{(total transfer counts)}}
\label{eqn_freq_reduction}
\end{multline}
\vspace{-0.4cm}
\begin{multline}
\text{network traffic demand reduction rate} = \\
\frac{\text{(total shared data size + total transfer size)}}{\text{(total transfer size)}}
\label{eqn_traffic_reduction}
\end{multline}

Network transfer frequency demand reduction rates on the entire regional cache, shown in Figure \ref{fig_net_reduction_ratio}a, reflect a similar pattern on each node in the regional cache. The network transfer frequency demand was reduced by a factor of 2.62 on the average over the study period.
Network traffic demand reduction rates on the entire regional cache, shown in Figure \ref{fig_net_reduction_ratio}b, also follow similar patterns to each node in the regional cache. Network traffic demand was reduced by a factor of 2.91 on average over the study period.

\begin{figure}[htb!]
\centering
\includegraphics[width=\columnwidth]{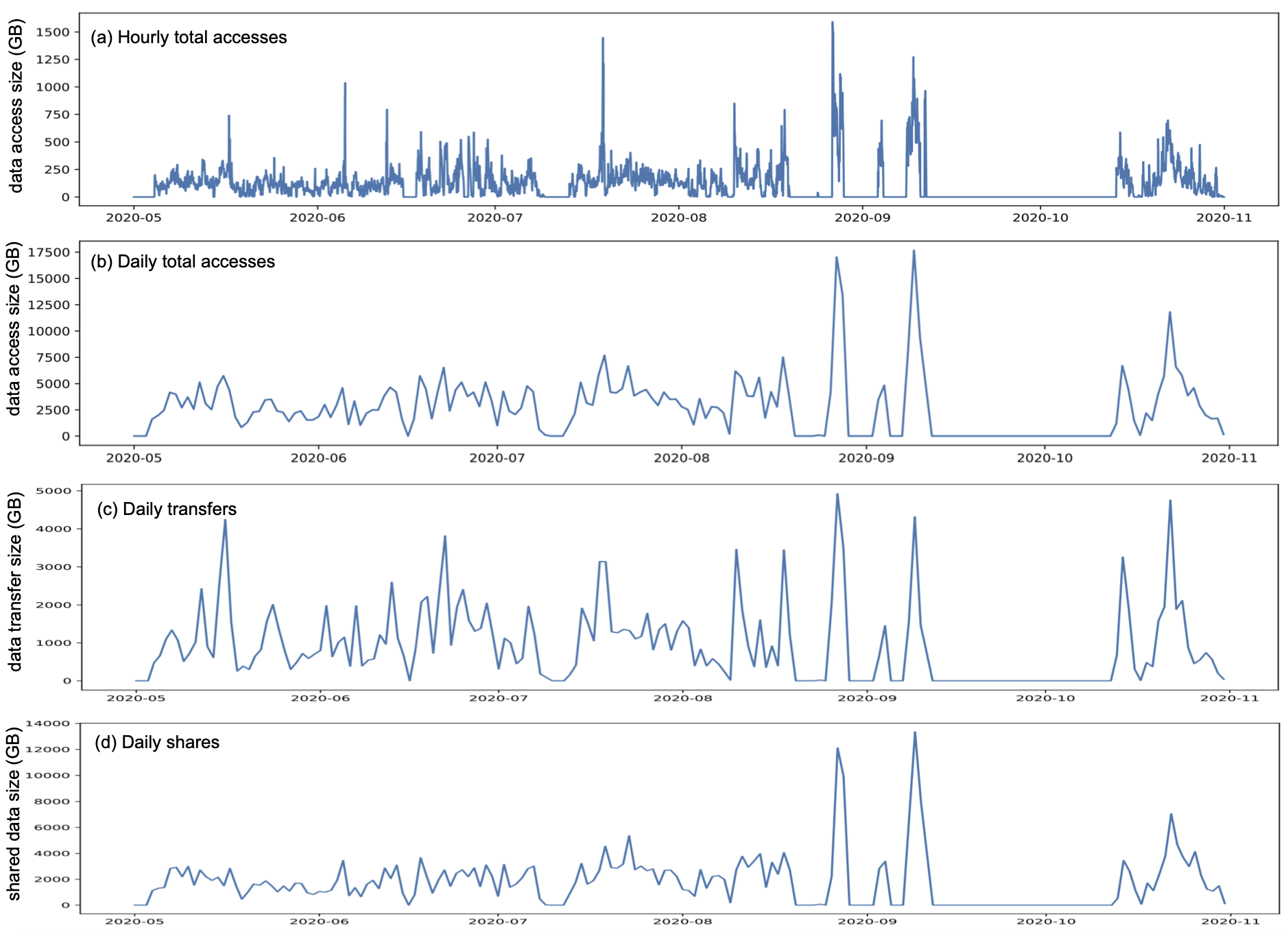}
\caption{
    Network traffic size on ESnet node; (a) hourly total data access size, (b) daily total data access size, (c) daily data transfer size, (d) daily shared data size
}
\label{fig_access_size}
\end{figure}

The network utilization of an individual cache node in the regional cache is expected to follow a similar pattern to the regional cache. 
Figure \ref{fig_access_size} shows the network traffic volume on the ESnet node during the study period. 

Figures \ref{fig_access_size}a and \ref{fig_access_size}b show the total data access size from applications to the ESnet node cache. These include data sizes from the first time and repeated time accesses to the files. 
Most of higher hourly volumes shown on the Figure \ref{fig_access_size}a are averaged out in the daily volumes in Figure \ref{fig_access_size}b.
Figure \ref{fig_access_size}c shows the data transfer sizes from the remote sites to the local ESnet node cache upon the first time application request for the data (cache misses). 
Figure \ref{fig_access_size}d shows the shared data sizes on the ESnet cache node which indicates the network traffic savings by the repeated data accesses (cache hits). 
Network traffic demand on the ESnet node was reduced by a factor of 2.92 on average over the study period.

In summary, the utilization of the temporary in-network data cache shows a reduction of redundant data transfers for scientific jobs, which consequently saves the network traffic volume. From our study on the ESnet cache node, the network demand is reduced by a factor of 2.92. 
From the total of 1,286,748 data accesses measured on the ESnet node from May 2020 to Oct. 2020, we observed the total of 490.831TB of client data accesses (first time reads and repeated reads), with 168.08TB of data transfers (from remote sites to the local cache), and 322.748TB of network traffic volume savings from the repeated shared data reads. 

%% file: related.tex
\section{Related work} \label{sec:related}
Delivering read-only scientific data from high energy physics experiments and climate modeling over Named Data Networking (NDN)~\cite{Zhang2014} was studied~\cite{Shannigrahi2015}. Although many benefits are offered in NDN by naming data entities rather than the network location of the data, further studies are needed with different scientific use cases to prove its worth. 
There have been many studies~\cite{vakali2003content, amadeo2016information, zeydan2016big, jiang2018cachalot, IBP, Wilkinson14storja, Syndicate, onedata, khan2019scispace} on Information Centric Networking (ICN), Content Delivery Network (CDN), cloud storages, and network caching. Our focus in this paper is on analyzing the data access patterns and network traffic savings from using the in-network caching method in the application of high energy physics.

%% file: conc.tex
\section{Conclusion} \label{sec:conclusion}
We analyzed measurements from the in-network caching infrastructure for a large scientific experiment and studied the cache utilization, network utilization, data access patterns, and impacts of the data caching nodes to the regional data repository.
About 65\% of data was shared in the regional cache on the average during the study period. 
The data access load is not linearly proportional to the caching storage size, but larger caching storage contributes to longer data holdings in the cache, resulting in less data access latency for applications as time goes on. 
Consequently, the network demand reduction rate increases as time goes on.
One ESnet cache node has a network traffic demand reduction rate of about 2.92 on average over the 6 month period. The regional cache as a whole also has the network traffic demand reduction rate of about 2.91 on the average over the study period, with a network transfer frequency demand reduction rate of about 2.62 on average during the study period.

Further work could be done to study data utilization, the correlation between network utilization and cache management, cache optimization, and application performance efficiency. In particular, some data have shown to be transferred multiple times over a period of time due to the limits on the caching space. We plan to explore how cache misses affect application performance and network performance, as well as how cache management would help those performances under certain conditions and policies.